\documentclass[conference]{IEEEtran}
\IEEEoverridecommandlockouts
\usepackage{amsmath,amssymb,amsfonts}
\usepackage{algorithmic}
\usepackage{graphicx}
\usepackage{textcomp}
\usepackage[table]{xcolor}

\usepackage{verbatim}
\usepackage{amsmath}
\usepackage{array}
\usepackage{dcolumn}
\usepackage{bm}
\usepackage{setspace}
\usepackage{amssymb}
\usepackage{braket}
\usepackage{graphicx}
\usepackage[caption=false]{subfig}

\usepackage{ragged2e}
\usepackage{multirow}

\usepackage{verbatim} 
\usepackage{mathtools}
\usepackage{enumitem}
\usepackage{multicol}

  {\list{}{\leftmargin=#1\rightmargin=#1}\item[]}%
  {\endlist}
  
\begin{document}
\bstctlcite{IEEEexample:BSTcontrol}
\author{\IEEEauthorblockN{
Charlene Yang 
}

\IEEEauthorblockA{
National Energy Research Scientific Computing Center (NERSC) \\
Lawrence Berkeley National Laboratory (LBNL) \\ Berkeley, California 94720, USA \\
}
\IEEEauthorblockN{
{
cjyang@lbl.gov 
}
}
}
\title{Hierarchical Roofline Analysis: \\How to Collect Data using Performance Tools \\on Intel CPUs and NVIDIA GPUs}

\maketitle

\begin{abstract}
This paper surveys a range of methods to collect necessary performance data on Intel CPUs and NVIDIA GPUs for hierarchical Roofline analysis. 
As of mid-2020, two vendor performance tools, Intel Advisor and NVIDIA Nsight Compute, have integrated Roofline analysis into their supported feature set. 
This paper fills the gap for when these tools are not available, or when users would like a more customized workflow for certain analysis.
Specifically, we will discuss how to use Intel Advisor, RRZE LIKWID, Intel SDE and Intel Amplifier on Intel architectures, and nvprof, Nsight Compute metrics, and Nsight Compute section files on NVIDIA architectures. 
These tools will be used to collect information for as many memory/cache levels in the memory hierarchy as possible in order to provide insights into application's data reuse and cache locality characteristics.

\end{abstract}


\section{Introduction}


The Roofline performance model \cite{CACM09_Roofline1} offers an insightful and intuitive way to extracte key computational characteristics for applications in high-performance computing (HPC).
Its capability to abstract away the complexity of modern memory hierarchies and guide performance analysis and optimization effort has gained its popularity in recent years.


Roofline is a throughput-oriented model centered around the interplay between computational capabilities, memory bandwidth, and data locality.  Data locality is the reuse of data once it is being loaded from the main memory, and it is commonly expressed as the arithmetic intensity (AI), ratio between the floating-point operations performed and the data moved (FLOPs/Byte).
The performance (GFLOP/s) is bound by the following two terms:
\begin{equation}
\mbox{GFLOP/s} \leq \mbox{min} \left\{ 
  \begin{array}{@{}l@{}}
    \mbox{Peak GFLOP/s}							\\
    \mbox{Peak GB/s} \times \mbox{Arithmetic Intensity}	
  \end{array}
\right.
\label{eq:roofline_naive}
\end{equation}




Conventionally, the Roofline model is focused on one level of the memory system, but this has been extended to the entire memory hierarchy in recent years, named the hierarchical Roofline model. 
The hierarchical Roofline helps understand cache reuse and data locality and provides additional insights into the efficiency of the application's utilization of the memory subsystem. 
The hierarchical Roofline has been integrated into Intel Advisor \cite{advisor,koskela2018novel}, and NVIDIA Nsight Compute \cite{ncu,nsight_compute}. 
Even though they should be the go-to methods for Roofline analysis, we would like to present in this paper a few other tools for the purpose of flexibility and generality. 

We will discuss the use of Intel Advisor \cite{advisor}, RRZE LIKWID \cite{likwid}, Intel SDE \cite{SDE} and Intel VTune \cite{VTune} on Intel CPUs, and nvprof \cite{nvprof}, Nsight Compute metrics, and Nsight Compute section files \cite{nsight_compute} on NVIDIA GPUs.
A mini-application will be used for demonstration and validation purpose, and it is extracted from the Material Science code BerkeleyGW \cite{BGW2} called General Plasmon Pole (GPP) \cite{examplescripts}. 
Architecture-wise, we will focus on the Intel Knights Landing (KNL) CPU and NVIDIA V100 GPU.


  

To facilitate the Roofline study, a range of other tools have sprung to life as well,
for example, the Empirical Roofline Toolkit (ERT) for more accurate machine characterization \cite{ert,yang2018empirical}, and \cite{nerscroofline,yang2018toward,madsen2020timemory,yang2019hierarchical,yang2019cug} for more streamlined data collection methods.
Other than tools development, there are many studies on the application of the Roofline model in traditional HPC \cite{Doerfler,yang2019hierarchical,yang2019cug,del2020accelerating,gayatri2018case} and Machine Learning \cite{yang2019hierarchical,yang2019cug,wang2020pmbs,wang2020dlonsc}, and extension and refinement of the model to related topics in HPC, such as instruction Roofline \cite{ding2019instruction}, time-based Roofline \cite{wang2020dlonsc}, Roofline scaling trajectory \cite{ibrahim2019performance}, performance portability analysis based on Roofline \cite{yang2018empirical}, and power and energy Roofline  \cite{powerroofline,alexpowerroofline}.


\section{Application and Machine Setup}

\subsection{Mini-Application General Plasmon Pole (GPP) \label{sec:gpp}}

The GPP mini-application~\cite{examplescripts} is extracted from the Material Science code BerkeleyGW~\cite{BGW2}, and it represents the work typically done on a single MPI rank.
It is written in C++, and parallelized with OpenMP on the CPU and CUDA on the GPU.
The computation involved this mini-app is tensor-contraction like, and several pre-calculated complex double precision arrays are multiplied and summed over certain dimensions and collapsed into a small vector. 
The problem used in this paper is a medium sized one, and it comprises of 512 electrons and 32768 plane wave basis elements.
The pseudo code for this mini-app is as follows.

{\footnotesize
\begin{verbatim}
  do band = 1, nbands   
   do igp = 1, ngpown  
    do ig = 1, ncouls   
     do iw = 1, nw     
       load wtilde_array(ig,igp)  
       load aqsntemp(ig,band)  
       load eps(ig,igp)  
       compute wdiff, delw, sch_array, ssx_array
       reduce on achtemp(iw), asxtemp(iw)
\end{verbatim}
}
The real code, job scripts and resulted are available at \cite{examplescripts}.

\subsection{Machine Setup\label{sec:machine}}

This study is conducted on the Cori supercomputer at the National Energy Research Scientific Computing Center (NERSC) at Lawrence Berkeley National Laboratory (LBNL).

Cori has three main partitions, Haswell, KNL and GPU, and this study has used its KNL partition \cite{nerscdocu1} and GPU chassis \cite{nerscdocu2}. 
Each KNL node is a single-socket Intel Xeon Phi Processor 7250 (Knights Landing) processor and has 68 physical cores.
There is 96 GB DDR4 memory and 16 GB MCDRAM (or HBM) per node, with the MCDRAM configured in `cache' mode by default.
The GPU chassis is deployed primarily for 
the NERSC Exascale Science Applications Program (NESAP).
It has 18 GPU nodes in total, and each node contains two 20-core Intel Xeon Gold 6148 Skylake CPUs, 384 GB DDR4 memory, 
and 8 NVIDIA V100 Volta GPUs.
Each GPU has 80 Streaming Multiprocessors (SMs), 16 GB HBM2 memory, and is connected to others in a `hybrid cube-mesh' topology.

\section{Methods and Results}

\subsection{Roofline Data Collection on Intel CPUs\label{sec:metho1}}

Intel Advisor \cite{advisor} provides the production quality, fully integrated hierarchical Roofline analysis on Intel CPUs, with very little user effort required. Compared to LIKWID \cite{likwid}, it has a higher profiling overhead due to the static instruction analysis and cache simulation. LIKWID \cite{likwid} is an open-source package developed at the Regional Computing Center Erlangen (RRZE) in Germany. It provides several 
`performance groups' for easier and more streamlined performance analysis, and in this paper, we have identified a few for the hierarchical Roolfine data collection.
LIKWID uses metrics that are based on micro-ops not instructions, and in some cases, it does not distinguish the different vector lengths, such as scalars, AVX-2/AVX-512 instructions, and masked/unmasked vector lanes.
This may cause certain inaccuracy and require extra care, however its low overhead has made it a very attractive option for large-scale application analysis.
To collect hierarchical Roofline data, another method is to use Intel SDE \cite{SDE} and VTune \cite{VTune}.
SDE has a very high profiling overhead but it provides the most accurate instruction count and it can produce L1 data movement information as well. 
On the other hand, VTune can be used to collect DDR/MCDRAM information to complement SDE.
In the following few subsections, we will detail the command lines used to collect Roofline data on KNL and the subsequent results.

\subsubsection{Intel Advisor}

\begin{figure}[h]
\centering
\makebox[0.5\textwidth][c]{\includegraphics[width=.5\textwidth]{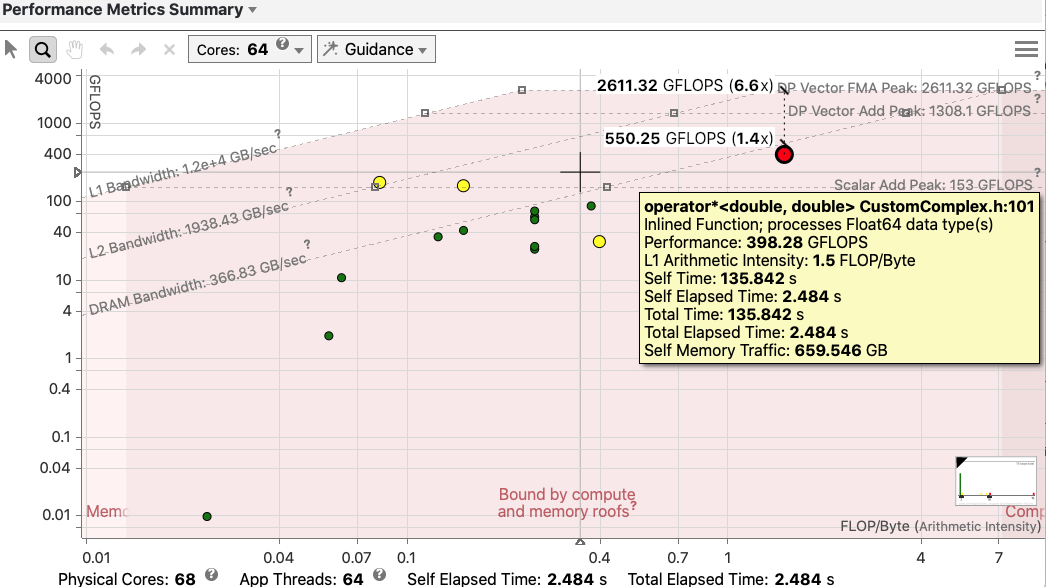}}
\caption{Roofline analysis of GPP on KNL using Advisor}
\label{fig:adv}
\end{figure}

Advisor can be invoked as follows for Roofline analysis, and Fig.~\ref{fig:adv} shows that in GPP, the most significant function takes 2s of `Self-Time' and produces 398 GFLOP/s double-precision performance on 64 OpenMP threads.
Advisor naturally provides details on the level of functions and loops, while the methods we will discuss below may require some code instrumentation in order to focus on certain code regions of interest. 

{\footnotesize
\begin{verbatim}
  module load advisor/2020
  advixe-cl --collect=roofline --project-dir=<dir>
        -- ./gpp 512 2 32768 20 0
\end{verbatim}
}

\subsubsection{RRZE LIKWID}

LIKWID \cite{likwid} is an open-source software package and here we use its `performance groups', FLOPS\_DP, HBM\_CACHE, L2 and DATA (for L1), for hierarchical Roofline data collection. 
Each of these groups contains a set of raw hardware counters and derived performance metrics, without user having to dive deep into the nitty-gritty micro-architecture specs and hardware counter details.
The following command can be used to profile with LIKWID, 

{\footnotesize
\begin{verbatim}
  module load likwid/4.3.0
  groups=('FLOPS_DP' 'HBM_CACHE' 'L2' 'DATA')
  for gs in ${groups[@]}
  do
    likwid-perfctr -c 0-271 -g $gs 
                   ./gpp 512 2 32768 20 0
  done
\end{verbatim}
}

\begin{figure}[h]
\centering
\makebox[0.5\textwidth][c]{\includegraphics[width=.5\textwidth]{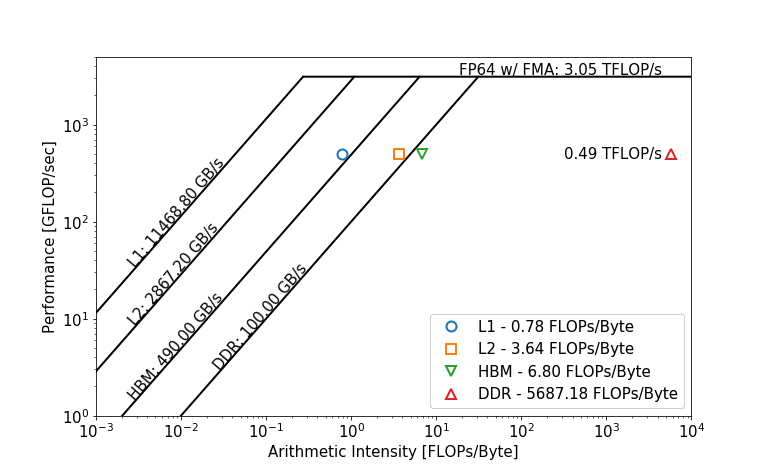}}
\caption{Roofline analysis of GPP on KNL using LIKWID}
\label{fig:likwid}
\end{figure}

The raw results for GPP are as follows, and Fig.~\ref{fig:likwid} shows that LIKWID produces a similar Roofline chart as Advisor, with close arithmetic intensity and performance.
The DDR-level arithmetic intensity is extremely high in Fig.~\ref{fig:likwid}, because the data set (1.5-2 GB) fits well into the HBM cache and there is little memory transaction between DDR and HBM.


{\footnotesize
\begin{verbatim}
  Time: 10.2243 secs
  GFLOPS: 5051.923 
  MCDRAM Bytes: 742.8158 GB
  DDR Bytes: 0.8883 GB
  L2 Bytes: 1387.739 GB
  L1 Bytes: 6456.799 GB
\end{verbatim}
}
%

\subsubsection{Intel VTune and Intel SDE}

\begin{figure}[h]
\centering
\makebox[0.5\textwidth][c]{\includegraphics[width=.5\textwidth]{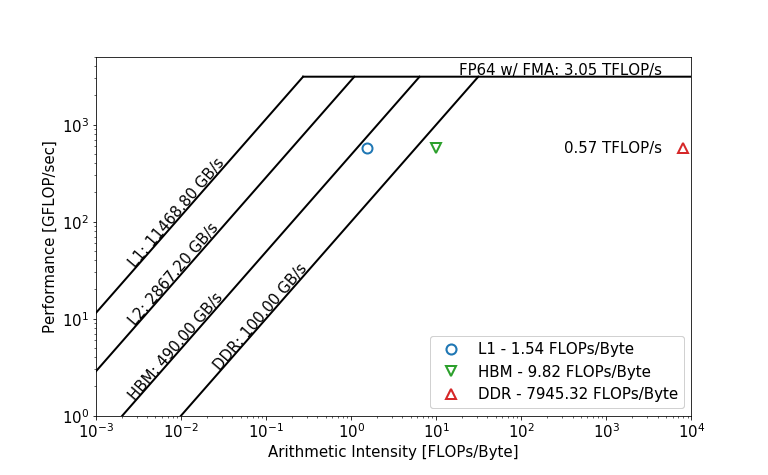}}
\caption{Roofline analysis of GPP on KNL using SDE and VTune}
\label{fig:sdevtune}
\end{figure}

This is a methodology developed a few years before the full integration of Roofline into Advisor, and may still present value to users who would like to investigate the underlying details.
In this case, the SDE tool can be used for collection of the FLOPs count and L1 data movement, while VTune can be used for uncore data movement collection.
The commands and results for GPP analysis in this paper are listed below, and Fig.~\ref{fig:sdevtune}  presents the combined data with a very high consistency with the results in Fig.~\ref{fig:likwid} (albeit the missing L2 data). 

{\footnotesize
\begin{verbatim}
  # commands for SDE
  sde64 -knl -d -iform 1 -omix result.sde 
        -global_region 
        -start_ssc_mark 111:repeat 
        -stop_ssc_mark 222:repeat 
        -- ./gpp 512 2 32768 20 0
\end{verbatim}
}
{\footnotesize
\begin{verbatim}
  # results from SDE
  GFLOPS: 5839.811
  L1 Bytes: 3795.623
\end{verbatim}
}
{\footnotesize
\begin{verbatim}
  # commands for VTune
  module load vtune/2020
  vtune -start-paused -r my-vtune.knl 
        -collect memory-access 
        -finalization-mode=none 
        -data-limit=0 
        -- ./gpp 512 2 32768 20 0
  vtune -report hw-events 
        -group-by=package 
        -r my-vtune.knl/ 
        -format csv -csv-delimiter comma  
        > advisor.html
\end{verbatim}
}
{\footnotesize
\begin{verbatim}
  # results from VTune
  DDR Bytes: 0.735
  MCDRAM Bytes: 594.562
\end{verbatim}
}

\subsection{Roofline Data Collection on NVIDIA GPUs\label{sec:metho2}}

On NVIDIA GPUs, an nvprof \cite{nvprof} based methodology was first proposed in  \cite{yang2019hierarchical}, then an Nsight Compute \cite{nsight_compute} metrics based one developed in \cite{wang2020pmbs,metho}. These methodologies require a dozen of metrics to be collected for hierarchical Roofline analysis, and could incur significant profiling overhead when the number of kernels in the code is high.
With nvprof phasing out in the developer toolchain, Nsight Compute has become the focus of the development of Roofline data collection methodology. 
A more simplified set of metrics are identified and validated in \cite{metho,yang20208}, and it has since been integrated into Nsight Compute 2020 (CUDA 11 release) \cite{ncu}.
The default Roofline feature shipped in Nsight Compute 2020 only includes the HBM level analysis, but it can be extended by using custom section files and/or job scripts such as \cite{metho,yang20208}, for hierarchical Roofline analysis.

\subsubsection{Custom Section Files in Nsight Compute 2020\label{sec:ncuprof}}
\begin{figure}[h]
\centering
\makebox[0.5\textwidth][c]{\includegraphics[width=.5\textwidth]{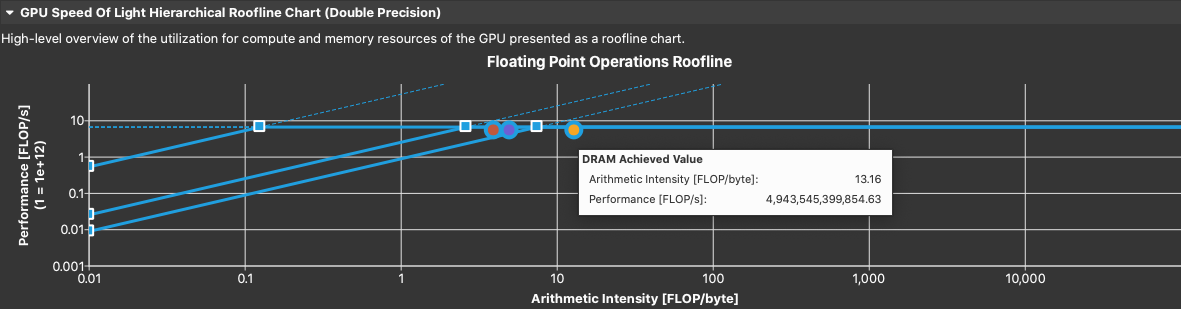}}
\caption{Roofline analysis of GPP on V100 using Nsight Compute 2020}
\label{fig:ncuprof}
\end{figure}

Nsight Compute uses Google Protocol Buffer messages for the section file, 
and it allows users to quickly create custom section files for their own tailored analysis. 
The following is an example in \cite{examplescripts} that can be used to collect the hierarchical double precision Roofline data for GPP, and its results are shown in Fig.~\ref{fig:ncuprof}.
The 13 FLOPs/Byte arithmetic intensity shows that this kernel has well entered the compute bound region on the HBM level, and particular attention should be paid to the utilization of compute resources such as threads and instructions, rather than the memory system.

{\footnotesize
\begin{verbatim}
  module load nsight-compute/2020.1.0
  ncu -k NumBandNgpown_kernel 
      -o ncu.prof 
      --section-folder ./ncu-section-files 
      --section 
      SpeedOfLight_HierarchicalDoubleRooflineChart 
      ./gpp 512 2 32768 20 0 
\end{verbatim}
}

\begin{table}[h]
\caption{{nvprof} metrics for Roofline data collection}
\label{tab:nvprof}
{\footnotesize
  \begin{tabular}{|c|c|}
  \hline
    & \textbf{Commands/Metrics}\\
  \hline\hline
\multirow{1}{*}{Time}
    & nvprof --print-gpu-summary ./gpp 512 2 32768 20 0\\
\hline\hline
\multirow{1}{*}{FP64 FLOPs}
    & nvprof --metrics flop\_count\_dp \\
    \hline
\multirow{1}{*}{ FP32 FLOPs }
    & flop\_count\_sp \\
\hline
\multirow{1}{*}{ FP16 FLOPs }
    & flop\_count\_hp \\
\hline
    Tensor Core & tensor\_precision\_fu\_utilization\\
\hline\hline
\multirow{3}{*}{L1 Cache}
         & gld\_transactions, gst\_transactions, atomic\_transactions\\ 
         & local\_load\_transactions, local\_store\_transactions \\
         & shared\_load\_transactions, shared\_store\_transactions \\
\hline
\multirow{1}{*}{L2 Cache}
    & l2\_read\_transactions, l2\_write\_transactions\\
\hline
\multirow{1}{*}{HBM}
    & dram\_read\_transactions, dram\_write\_transactions\\
\hline
  \end{tabular}
 }
\end{table}

\subsubsection{The nvprof Profiler}

\begin{figure}[h]
\centering
\makebox[0.5\textwidth][c]{\includegraphics[width=.5\textwidth]{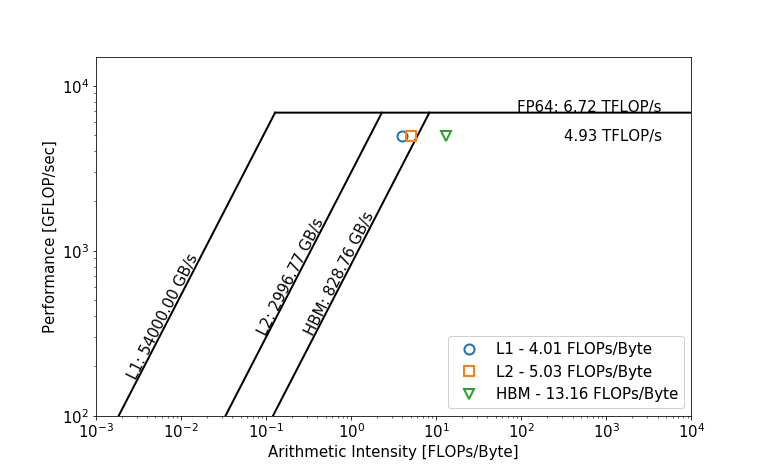}}
\caption{Roofline analysis of GPP on V100 using nvprof metrics}
\label{fig:nvprof}
\end{figure}

Many developers started their GPU optimization with the nvprof profiler and our initial Roofline methodology also starts with the metrics in nvprof.
Tab.~\ref{tab:nvprof} lists a set of metrics that can be used for hierarchical Roofline analysis and they are put in three categories, runtime, FLOPs count, and data movement (in bytes) between different memory/cache levels.
These metrics are based on CUPTI and can be mapped to the PerfWorks framework in Nsight Compute through \cite{metricsmapping}, with certain validation.
The following command has been used for the GPP data collection and the results are in Fig.~\ref{fig:nvprof}, with a very similar set of arithmetic intensities on L1, L2 and HBM levels, and GFLOP/s performance to those in Fig.~\ref{fig:ncuprof}.

{\footnotesize
\begin{verbatim}
  module load cuda/10.2.89
  metrics='fp_count_dp,...' # see Tab. I
  nvprof --kernels NumBandNgpown_kernel
         --metrics $metrics 
         ./gpp 512 2 32768 20 0
\end{verbatim}
}


\subsubsection{Metrics in Nsight Compute 2019}

\begin{figure}[h]
\centering
\makebox[0.5\textwidth][c]{\includegraphics[width=.5\textwidth]{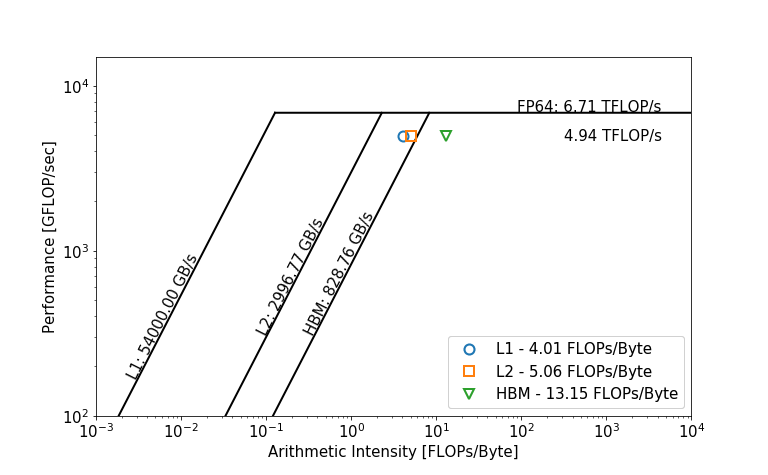}}
\caption{Roofline analysis of GPP on V100 using Nsight Compute 2019 metrics}
\label{fig:ncu10}
\end{figure}

As nvprof phases out, we have developed a data collection methodology based on Nsight Compute 2019.
These metrics as listed in Tab.~\ref{tab:ncu10} are more detailed than those in nvprof, and they produce comparable results as seen in Fig.~\ref{fig:ncu10} and Fig.~\ref{fig:ncuprof}.
The commands used to collect Roofline data for GPP are as follows.

{\footnotesize
\begin{verbatim}
  module load cuda/10.2.89
  metrics='sm__cycles_elapsed.avg,...' # see Tab. II
  nv-nsight-cu-cli  -k NumBandNgpown_kernel
                    --metrics $metrics 
                    ./gpp 512 2 32768 20 0 
\end{verbatim}
}

\begin{table}[h]
\caption{Nsight Compute 2019 metrics for Roofline data collection}
\label{tab:ncu10}
{\footnotesize
  \begin{tabular}{|@{\;}c@{\;\;}|@{\;\;}c@{\;}|}
  \hline
    & \textbf{Metrics}\\
  \hline\hline
\multirow{2}{*}{Time}
    & sm\_\_cycles\_elapsed.avg\\
    & sm\_\_cycles\_elapsed.avg.per\_second\\
\hline\hline
\multirow{3}{*}{FP64 FLOPs}
    & sm\_\_sass\_thread\_inst\_executed\_op\_dadd\_pred\_on.sum\\
    & sm\_\_sass\_thread\_inst\_executed\_op\_dmul\_pred\_on.sum\\
    & sm\_\_sass\_thread\_inst\_executed\_op\_dfma\_pred\_on.sum\\
    \hline
\multirow{3}{*}{FP32 FLOPs}
    & sm\_\_sass\_thread\_inst\_executed\_op\_fadd\_pred\_on.sum\\
    & sm\_\_sass\_thread\_inst\_executed\_op\_fmul\_pred\_on.sum\\
    & sm\_\_sass\_thread\_inst\_executed\_op\_ffma\_pred\_on.sum \\
\hline
\multirow{3}{*}{FP16 FLOPs}
    & sm\_\_sass\_thread\_inst\_executed\_op\_hadd\_pred\_on.sum\\
    & sm\_\_sass\_thread\_inst\_executed\_op\_hmul\_pred\_on.sum\\
    & sm\_\_sass\_thread\_inst\_executed\_op\_hfma\_pred\_on.sum\\
\hline
    Tensor Core & sm\_\_inst\_executed\_pipe\_tensor.sum\\
\hline\hline
\multirow{10}{*}{L1 Cache}
    & l1tex\_\_t\_sectors\_pipe\_lsu\_mem\_global\_op\_ld.sum\\
    & l1tex\_\_t\_bytes\_pipe\_lsu\_mem\_global\_op\_st.sum\\
    & l1tex\_\_t\_set\_accesses\_pipe\_lsu\_mem\_global\_op\_atom.sum\\
    & l1tex\_\_t\_set\_accesses\_pipe\_lsu\_mem\_global\_op\_red.sum\\
    & l1tex\_\_t\_set\_accesses\_pipe\_tex\_mem\_surface\_op\_atom.sum\\
    & l1tex\_\_t\_set\_accesses\_pipe\_tex\_mem\_surface\_op\_red.sum\\
    & l1tex\_\_t\_sectors\_pipe\_lsu\_mem\_local\_op\_ld.sum\\
    & l1tex\_\_t\_sectors\_pipe\_lsu\_mem\_local\_op\_st.sum\\
    & l1tex\_\_data\_pipe\_lsu\_wavefronts\_mem\_shared\_op\_ld.sum\\
    & l1tex\_\_data\_pipe\_lsu\_wavefronts\_mem\_shared\_op\_st.sum\\
\hline
\multirow{4}{*}{L2 Cache}
    & lts\_\_t\_sectors\_op\_read.sum\\
    & lts\_\_t\_sectors\_op\_write.sum\\
    & lts\_\_t\_sectors\_op\_atom.sum\\
    & lts\_\_t\_sectors\_op\_red.sum\\
\hline
\multirow{2}{*}{HBM}
    & dram\_\_sectors\_read.sum\\
    & dram\_\_sectors\_write.sum\\
\hline
  \end{tabular}
  }
\end{table}

\begin{table}[h]
\caption{Nsight Compute 2020 metrics for Roofline data collection}
\label{tab:ncu11}
{\footnotesize
  \begin{tabular}{|c|c|}
  \hline
    & \textbf{Commands/Metrics}\\
  \hline\hline
\multirow{2}{*}{Time}
    & sm\_\_cycles\_elapsed.avg \\ 
    & sm\_\_cycles\_elapsed.avg.per\_second \\
\hline\hline
\multirow{3}{*}{FP64 FLOPs}
& sm\_\_sass\_thread\_inst\_executed\_op\_dadd\_pred\_on.sum \\
& sm\_\_sass\_thread\_inst\_executed\_op\_dfma\_pred\_on.sum \\
& sm\_\_sass\_thread\_inst\_executed\_op\_dmul\_pred\_on.sum \\
    \hline
\multirow{3}{*}{ FP32 FLOPs }
    & sm\_\_sass\_thread\_inst\_executed\_op\_fadd\_pred\_on.sum \\
    & sm\_\_sass\_thread\_inst\_executed\_op\_ffma\_pred\_on.sum \\
    & sm\_\_sass\_thread\_inst\_executed\_op\_fmul\_pred\_on.sum \\
\hline
\multirow{3}{*}{ FP16 FLOPs }
    & sm\_\_sass\_thread\_inst\_executed\_op\_hadd\_pred\_on.sum \\
    & sm\_\_sass\_thread\_inst\_executed\_op\_hfma\_pred\_on.sum \\
    & sm\_\_sass\_thread\_inst\_executed\_op\_hmul\_pred\_on.sum \\
\hline
     Tensor Core  & \multirow{1}{*}{sm\_\_inst\_executed\_pipe\_tensor.sum}\\
\hline\hline
\multirow{1}{*}{L1 Cache}
         & l1tex\_\_t\_bytes.sum \\
\hline
\multirow{1}{*}{L2 Cache}
    & lts\_\_t\_bytes.sum \\
\hline
\multirow{1}{*}{HBM}
    & dram\_\_bytes.sum\\
\hline
  \end{tabular}
 }
\end{table}

\subsubsection{Metrics in Nsight Compute 2020}

As Nsight Compute evolves over time, we have also developed a more simplified data collection methodology with fewer metrics to collect (please see Tab.~\ref{tab:ncu11}).
These metrics are equivalent to the ones used in section files in \ref{sec:ncuprof}, and scripts based on them \cite{metho} can be used for easier integration with users' other job submission workflows, and for more customized Roofline presentation (using Matplotlib).
The commands we used to collect Roofline information for GPP in this paper are as follows.

 
{\footnotesize
\begin{verbatim}
  module load nsight-compute/2020.1.0
  metrics='sm__cycles_elapsed.avg,...'
  ncu -k NumBandNgpown_kernel
      --metrics $metrics 
      ./gpp 512 2 32768 20 0  
\end{verbatim}
}

\begin{figure}[h]
\centering
\makebox[0.5\textwidth][c]{\includegraphics[width=.5\textwidth]{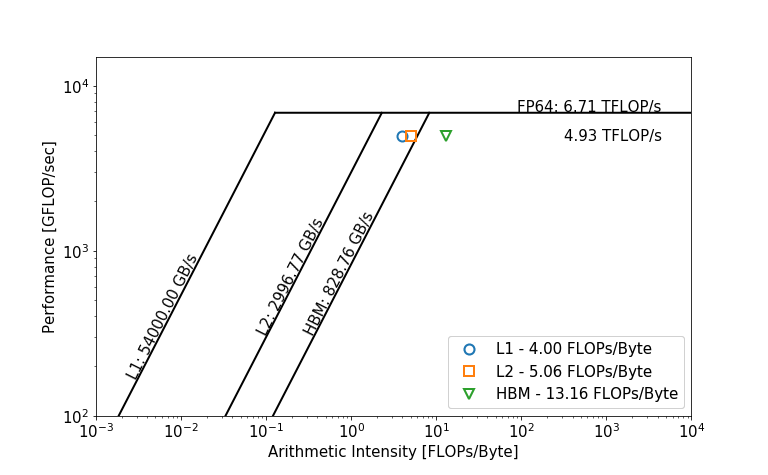}}
\caption{Roofline analysis of GPP on V100 using Nsight Compute 2020 metrics}
\label{fig:ncu11}
\end{figure}

Fig.~\ref{fig:ncu11} shows that this methodology produces consistent results as in previous subsections, with very 
marginal difference on the arithmetic intensity and GFLOP/s throughput.



\section{Summary}

In this paper, we have presented a range of methods using a variety of performance tools to collect hierarchical data for Roofline analysis. 
Even though the Roofline model has been integrated into production tools such as Intel Advisor and NVIDIA Nsight Compute, we still expect that this paper fills the gaps for developers who do not have access to those tools, or who would like to investigate the underlying details.
It would serve the purpose of flexibility and generality in the Roofline data collection space.


\bibliographystyle{IEEEtran}
\bibliography{IEEEabrv,references}

\end{document}